\documentstyle[12pt,epsf]{article} \hoffset -0.5in \textwidth 5.5in \textheight
8.5in  \parskip 7pt 
\newskip\humongous \humongous=0pt plus 1000pt minus 1000pt
\def\caja{\mathsurround=0pt}
\def\eqalign#1{\,\vcenter{\openup1\jot \caja
        \ialign{\strut \hfil$\displaystyle{##}$&$
        \displaystyle{{}##}$\hfil\crcr#1\crcr}}\,}
\newif\ifdtup

\def\eqright #1\cr{\noalign{\hfill$\displaystyle{{}#1}$}}
\def\eqleft #1\cr{\noalign{\noindent$\displaystyle{{}#1}$\hfill}}

\def\oldreffmt#1{\rlap{[#1]} \hbox to 2\parindent{}}

\def\figfmt#1{\rlap{Figure {#1}} \hbox to 1in{}}

\def\sectioneq{\def\theequation{\thesection.\arabic{equation}}{\let
\holdsection=\section\def\section{\setcounter{equation}{0}\holdsection}}}%

\newcounter{holdequation}

\def\begineq #1\endeq{$$ \refstepcounter{equation}\eqalign{#1}\eqno
	(\theequation) $$}
\def\contlimit{\,{\hbox{$\longrightarrow$}\kern-1.8em\lower1ex
\hbox{${\scriptstyle (a\rightarrow0)}$}}\,}
\def\centeron#1#2{{\setbox0=\hbox{#1}\setbox1=\hbox{#2}\ifdim
\wd1>\wd0\kern.5\wd1\kern-.5\wd0\fi
\copy0\kern-.5\wd0\kern-.5\wd1\copy1\ifdim\wd0>\wd1
\kern.5\wd0\kern-.5\wd1\fi}}
\def\centerover#1#2{\centeron{#1}{\setbox0=\hbox{#1}\setbox
1=\hbox{#2}\raise\ht0\hbox{\raise\dp1\hbox{\copy1}}}}
\def\centerunder#1#2{\centeron{#1}{\setbox0=\hbox{#1}\setbox
1=\hbox{#2}\lower\dp0\hbox{\lower\ht1\hbox{\copy1}}}}
\def\lsim{\;\centeron{\raise.35ex\hbox{$<$}}{\lower.65ex\hbox
{$\sim$}}\;}
\def\gsim{\;\centeron{\raise.35ex\hbox{$>$}}{\lower.65ex\hbox
{$\sim$}}\;}
\def\st#1{\centeron{$#1$}{$/$}}

\def\super#1{\ifmmode \hbox{\textsuper{#1}}\else\textsuper{#1}\fi}
\def\textsuper#1{\newcount\holdspacefactor\holdspacefactor=\spacefactor
$^{#1}$\spacefactor=\holdspacefactor}

\def\getcite#1,{\advance\citenumber by1
\def\getcitearg{#1}\def\lastarg{@}
\ifnum\citenumber=1
\ref{#1}\let\next=\getcite\else\ifx\getcitearg\lastarg\let\next=\relax
\else ,\ref{#1}\let\next=\getcite\fi\fi\next}

\def\pom{{\rm P\kern -0.53em\llap I\,}}
\def\spom{{\rm P\kern -0.36em\llap \small I\,}}
\def\sspom{{\rm P\kern -0.33em\llap \footnotesize I\,}}

\relax

\def\begineq #1\endeq{$$ \refstepcounter{equation}\eqalign{#1}\eqno
	(\theequation) $$}
\def\contlimit{\,{\hbox{$\longrightarrow$}\kern-1.8em\lower1ex
\hbox{${\scriptstyle (a\rightarrow0)}$}}\,}
\def\centeron#1#2{{\setbox0=\hbox{#1}\setbox1=\hbox{#2}\ifdim
\wd1>\wd0\kern.5\wd1\kern-.5\wd0\fi
\copy0\kern-.5\wd0\kern-.5\wd1\copy1\ifdim\wd0>\wd1
\kern.5\wd0\kern-.5\wd1\fi}}
\def\centerover#1#2{\centeron{#1}{\setbox0=\hbox{#1}\setbox
1=\hbox{#2}\raise\ht0\hbox{\raise\dp1\hbox{\copy1}}}}
\def\centerunder#1#2{\centeron{#1}{\setbox0=\hbox{#1}\setbox
1=\hbox{#2}\lower\dp0\hbox{\lower\ht1\hbox{\copy1}}}}
\def\lsim{\;\centeron{\raise.35ex\hbox{$<$}}{\lower.65ex\hbox
{$\sim$}}\;}
\def\gsim{\;\centeron{\raise.35ex\hbox{$>$}}{\lower.65ex\hbox
{$\sim$}}\;}
\def\st#1{\centeron{$#1$}{$/$}}

\def\super#1{\ifmmode \hbox{\textsuper{#1}}\else\textsuper{#1}\fi}
\def\textsuper#1{\newcount\holdspacefactor\holdspacefactor=\spacefactor
$^{#1}$\spacefactor=\holdspacefactor}

\def\getcite#1,{\advance\citenumber by1
\ifnum\citenumber=1
\ref{#1}\let\next=\getcite\else\ifx#1@\let\next=\relax
\else ,\ref{#1}\let\next=\getcite\fi\fi\next}

\def\upon #1/#2 {{\textstyle{#1\over #2}}}
\relax
\renewcommand{\thefootnote}{\fnsymbol{footnote}} 

\def\mainhead#1{\setcounter{equation}{0}\addtocounter{section}{1}
  \vbox{\begin{center}\large\bf #1\end{center}}\nobreak\par}
\sectioneq

\def\til#1{\centeron{\hbox{$#1$}}{\lower 2ex\hbox{$\char'176$}}}
\def\tild#1{\centeron{\hbox{$\,#1$}}{\lower 2.5ex\hbox{$\char'176$}}}
\def\sumtil{\centeron{\hbox{$\displaystyle\sum$}}{\lower
-1.5ex\hbox{$\widetilde{\phantom{xx}}$}}}

\def\pom{{\rm P\kern -0.53em\llap I\,}}
\def\spom{{\rm P\kern -0.36em\llap \small I\,}}
\def\sspom{{\rm P\kern -0.33em\llap \footnotesize I\,}}

\newcommand{\bit}{\begin{itemize}}
\newcommand{\eit}{\end{itemize}}

\newcommand{\beq}{\begin{equation}}
\newcommand{\eeq}{\end{equation}}
\newcommand{\beqa}{\begin{eqnarray}}
\newcommand{\eeqa}{\end{eqnarray}}

\begin{document} 

\rightline{\vbox{\halign{&#\hfil\cr
&ANL-HEP-CP-97-25 \cr
&\today\cr}}} 
\vspace{1.25in} 

\begin{center} 
 
{\large\bf 
THE POMERON BEYOND BFKL }\footnote{Work 
supported by the U.S.
Department of Energy, Division of High Energy Physics, \newline Contracts
W-31-109-ENG-38 and DEFG05-86-ER-40272} 
\medskip

Alan. R. White\footnote{arw@hep.anl.gov }

\vskip 0.6cm

\centerline{High Energy Physics Division}
\centerline{Argonne National Laboratory}
\centerline{9700 South Cass, Il 60439, USA.}
\vspace{0.5cm}

\end{center}

\begin{abstract} 

Conformally invariant reggeon interactions derived from $t$-channel 
unitarity are discussed and progress towards understanding the 
``physical Pomeron'', via massless quark reggeon interactions, is briefly 
outlined.

\end{abstract} 

\vspace{2in}
\begin{center}
Presented at the 5th International Workshop on Deep Inelastic 
Scattering and QCD, Chicago, Illinois, USA, April 14-18, 1997
\end{center}

\renewcommand{\thefootnote}{\arabic{footnote}} 


\newpage

\mainhead{1. INTRODUCTION} 

I will discuss two topics that
go beyond the BFKL Pomeron in QCD. The first is the 
derivation of conformally invariant reggeon interactions from $t$-channel 
unitarity. In particular $ln^4 [ \rho_{11'}\rho_{22'}/\rho_{12'}\rho_{1'2}]$
should appear\cite{cww}, in some approximation, in the NLO BFKL kernel and
higher powers of the same logarithm can be expected in higher orders. 

Secondly I will briefly outline my work in progress\cite{arw1}
that is aimed at understanding the ``physical Pomeron'' via the reggeon
interactions due to massless quarks. I discuss how triple-Regge vertices
contain an infra-red anomaly which produces reggeon diagram infra-red
divergences at zero quark mass when the SU(3) gauge symmetry is broken to
SU(2). The resulting divergent amplitudes contain a ``single gluon''
SuperCritical Pomeron and, I hope to prove, a hadron spectrum with
confinement and chiral symmetry breaking. I begin by briefly listing the
elements of the formalism I use\cite{arw2}. 

\mainhead {2. ANALYTIC MULTI-REGGE THEORY} 

The key ingredients are
\newline \parbox{3.8in}{ 	
{\it i)  Angular Variables } - For an N-particle amplitude
\newline $ M_N(P_1,\ldots,P_N) \equiv
M_N\left(t_1,\ldots,t_{N-3},g_1,\ldots,g_{N-3}\right)$
where $t_j=Q_j^2$ and $g_j \in $ the little group of $Q_j$, i.e. 
$g_j \in$ SO(3) or 
$g_j \in$ SO(2,1) for $t_j 
{\raisebox{1mm}{\centerunder{$\scriptscriptstyle 
>$}{$\scriptscriptstyle <$}}}~ 0.$ 
There are 
\newline (N-3) $t_i$ variables, 
(N-3) $z_j (\equiv \cos\theta_j)$ variables 
and 
\newline (N-4) $u_{jk} (\equiv e^{i(\mu_j - \nu_k)})$ variables.
}
\parbox{1.6in}{
\begin{center}
\leavevmode
\epsfxsize=1.4in
\epsffile{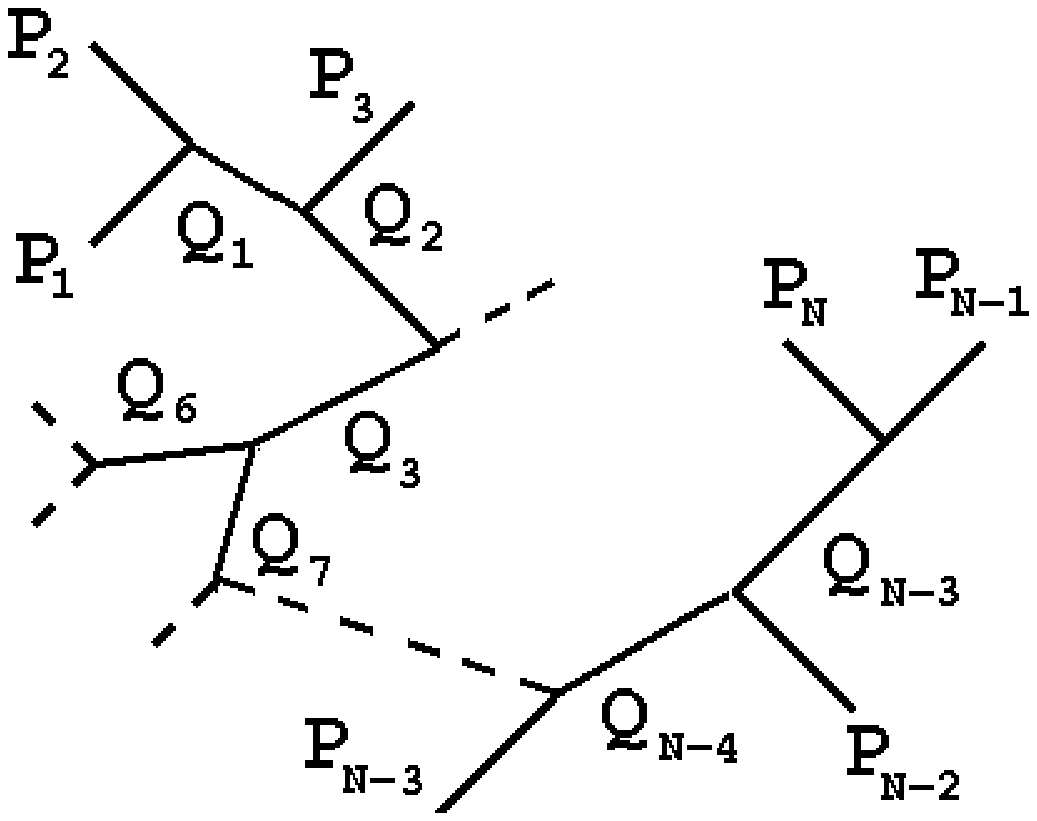}
\end{center}
}

\noindent ii) {\it Multi-Regge Limits } -  
all $z_j \to \infty$; {\it Helicity-Pole Limits} - some $u_{jk} \to \infty$.
\vspace{0.1in}
\newline iii) {\it Partial-wave Expansions} - 
$ f(g)=\sum^\infty_{J=0}\,\sum_{|n|,|n'|<J}D^J_{nn'}(g)a_{J nn'}$, 
$g \in$ SO(3) 
\newline $~~~~~ \to ~~M_N(\til{t},\til{g})=\sum_{\til{J},\til{n},\til{n'}}
\prod_i~D^{J_i}_{n_in_i'}(g_i)~ a_{\til{J},\til{n},\til{n'}} (\til{t})
$

\noindent iv) {\it Asymptotic Dispersion Relations} -
$M_N(p_1,..p_N)=\sum_{C}M_N^C(p_1,.. p_N)+M^0$ 
$  M_N^C(p_1,.. p_N)={1\over (2\pi i)^{N-3}}
\int \frac{ ds'_1\ldots ds'_{N-3}\Delta^C(
..t_i.,..u_{jk}.,..s'_i.)}
{(s'_1-s_1)(s'_2-s_2)\ldots (s'_{N-3}-s_{N-3})}
~~~~$
and $\sum_{C}$ is over all sets of (N-3) Regge limit asymptotic cuts.
\vspace{0.1in}
\newline v) {\it Sommerfeld-Watson Representations 
of Spectral Components} e.g.
\vspace{0.1in}
\newline $M^C_4={1\over 8}\sum_{{\scriptstyle N_1, N_2}} \int
{dn_2  dn_1 dJ_1 ~u_2^{n_2} u_1^{n_1}
d^{J_1}_{0,n_1}(z_1)
d^{n_1+N_2}_{n_1,n_2}(z_2)d^{n_2+N_3}_{n_2,0}(z_3)
\over
\sin\pi n_2\sin\pi(n_1-n_2)\sin\pi(J_1-n_1)}~a^C_{N_2N_3}(J_1,n_1,n_2,
\til{t})
$
\newline from which the form of multi-Regge 
behaviour in any limit can be extracted. 
\vspace{0.1in}
\newline vi) {\it $t$-channel Unitarity in the $J$-plane}
~~~Multiparticle unitarity can be partial-wave projected, diagonalized, and 
continued to complex $J$ in the form 
\vspace{0.1in}
\newline $ ~~~~~~~a^+_J - a^-_J= i\int d\rho \sum_{\til{N}} 
\int {dn_1 dn_2  \over 
sin\pi(J-n_1-n_2) }\int {dn_3 dn_4 \over sin\pi(n_1 -n_3 -n_4)} \cdots 
a^+_{J\til{N}
\til{n}}a^-_{J\til{N}\til{n}} 
$
\newline Regge poles at $n_i=\alpha_i$, together with ``nonsense poles'' at 
$J= n_1 +n_2 -1, n_1=n_3 + n_4 -1 , ...$ generate Regge cuts. 
The $J$-plane regge cut discontinuity due to $M$
Regge poles $\til{\alpha}= (\alpha_1, \alpha_2, \cdots \alpha_M)$ 
\newline $ ~~~~~
\centerunder{disc}{\raisebox{-3mm}{$\scriptstyle J=\alpha_M(t)$}}~~ 
a_{\til{N} \til{n}}(J) 
~=~ {\xi}_{M} \int d\hat\rho~
a_{\til{\alpha}}(J^+)
a_{\til{\alpha}}(J^-)
{\delta\left(J-1-\sum^M_{k=1}
(\alpha_k-1)\right)\over \sin{\pi\over 2}(\alpha_1-\tau'_1)\ldots\sin{\pi\over
2}(\alpha_M-\tau'_M) }
$
\newline is referred to as {\it reggeon unitarity}. Because the gluon
``reggeizes'', reggeon unitarity is a strong constraint on 
multigluon exchange amplitudes. 

\mainhead{3. NLO CONFORMAL SYMMETRY. }

The $J$-plane unitarity equations can also be used\cite{cw}to study 
the $t$-channel 
thresholds of reggeon (reggeized gluon) interactions due to nonsense gluon 
(particle) states. The gauge group is inserted via Regge pole vertices.
Known leading log results can be easily rederived. In particular
gluon reggeization is due to two-gluon nonsense  states (\epsfxsize=1.8in 
\epsffile{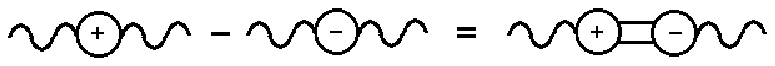})
while the $O(g^2)$ BFKL kernel is given by the three-gluon nonsense state
in the two-reggeon interaction ($\raisebox{-1mm}{\epsfxsize=0.5in
\epsffile{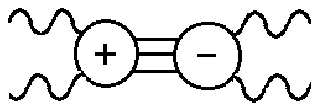}}$). 
The four-gluon nonsense state gives a NLO $~O(g^4)$ contribution $K^{(4)}$
to the BFKL kernel 
($\raisebox{-1mm}{\epsfxsize=0.5in \epsffile{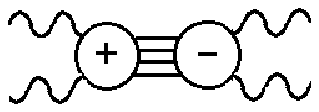}}$).

The phase-space integration $\int d\rho $ gives two-dimensional $k_{\perp}$ 
integrals via \newline $\int dt_1 dt_2 \lambda^{-1/2}(t,t_1,t_2) 
=~2\int d^2 k_{\perp}$. Using transverse momentum diagrams\cite{cw}
\newline \parbox{0.6in}{~\newline 
BFKL 
\newline kernel 
\newline 
~\newline 
$K^{(4)}$ 
\newline kernel 
\newline ~
\newline ~
}
\parbox{.5in}{
\begin{center}
~\newline 
$\leftrightarrow$
\newline ~
\newline ~
$\leftrightarrow$
\newline ~
\newline ~
\end{center}
}
\parbox{2.5in}{
\leavevmode
\epsfxsize=2.2in
\epsffile{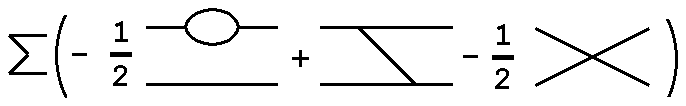}
\vspace{.1in}
\epsfxsize=2.2in
\epsffile{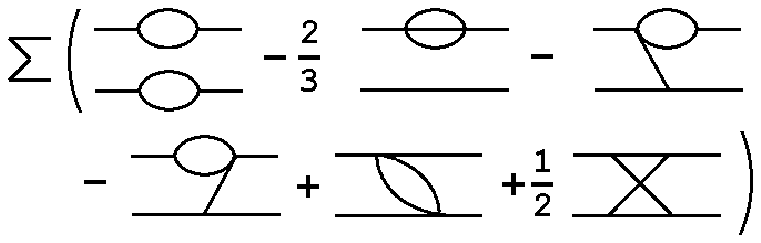}
}
\parbox{1.6in}{
Since the unitarity 
\newline analysis
has no scale, 
\newline if the results obtained are infra-red finite they 
are automatically scale-invariant.}
\newline Using eigenfunctions $\phi_{\nu,n}(k) = |k|^{\nu}e^{i{n \over 
2}\theta}$, the spectrum of $K^{(4)}$ has the form\cite{cw}
\newline $ {\cal E}(\nu,n)~=~{1 \over \pi} [\chi(\nu,n)]^2~-~\Lambda(\nu,n) $
where $\chi(\nu,n)$ are the eigenvalues of the BFKL kernel and 
$ \Lambda(\nu,n) =-{1 \over 4\pi}
\biggl(\beta'\bigl({|n| + 1\over 2} + 
i\nu\bigr)
+\beta'\bigl({|n| + 1 \over 2} -i\nu\bigr)\biggr)
$
with 
\newline $\beta(x)=\int^1_0 dy~y^{x -1}[1+y]^{-1}$. 
The holomorphic separability of $\Lambda(\nu,n)$, together with
BFKL conformal invariance, suggests that the impact parameter space 
kernel $\tilde{K}^{(4)}(\rho_1,\rho_2,\rho_{1'},\rho_{2'})$ should have a
conformal invariance property. 

Gauge invariance allows terms independent of any 
of $\rho_1,\rho_2,\rho_{1'},\rho_{2'}$ to be added to $\tilde{K}^{(4)}$. 
Using
$\int d^2 k ~ e^{i k.\rho} / (k^2 + m^2) 
= ~ - ln[m|\rho|/2] +\psi(1)+O(m)$
the two gluon 
propagator $ {1 \over J-1 }
~{\delta^2(k_1-k_{1'}) \over k_1^2}
{\delta^2(k_2-k_{2'}) \over k_2^2}$ gives the $\rho$-space analog
${4 \over (2\pi )^4(J-1)} \ln|\rho_{11'}|\ln|\rho_{22'}|$ 
(where $\rho_{11'} = \rho_1-\rho_{1'}$ etc.) which, after symmetrizing under $1 
\leftrightarrow 2$, is equivalent to $~{4 \over (2\pi )^4(J-1)} \ln^2 R $,
with $R= \left| \rho_{11'}\rho_{22'} / \rho_{12'}\rho_{1'2}\right|$. 
Evaluating the $k_{\perp}$ diagrams of $K^{(4)}$ gives that, up to terms 
that can be dropped because of
gauge invariance, each diagram also has\cite{cww} a $\rho$ space
analog with $\ln |\rho_{ij'}|$ ``propagators'', e.g.
~$\raisebox{-1mm}{\epsfxsize=0.4in \epsffile{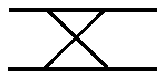}}
\to ~ln|\rho_{12'}|  ln|\rho_{21'}|  ln|\rho_{11'}|  
ln|\rho_{22'}|~+~....$  The sum of all diagrams gives the remarkably
simple, manifestly conformally invariant, representation\cite{cww}
\newline $~~~~~~~~~~~~~~~~~~~~~~~~~~~~~~~~~~~~~~~~~~~~
\tilde{K}^{(4)} \leftrightarrow  ~{1 \over 24} \ln^4 R $
\newline This representation was initially found 
via the Feynman diagram NNLO calculation of the large rapidity scattering of
two virtual photons. 
\vspace{0.1in} 
\newline $~~~$ Note that $ln^3R$ contains the 
diagrams $\raisebox{-2mm}{\epsfxsize=1.3in \epsffile{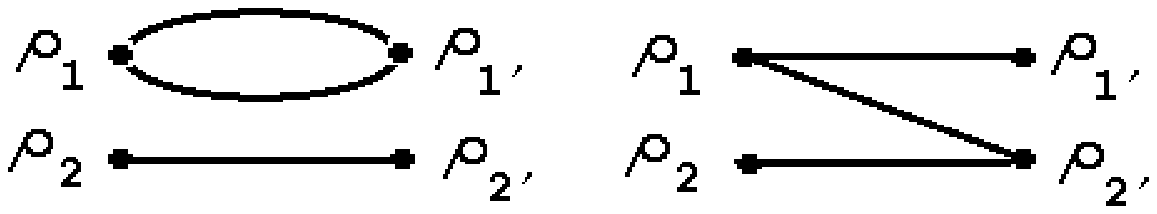}}$ that 
are naturally associated with the BFKL kernel. 
In fact the spectrum 
result for $K^{(4)}$ implies that (formally) we can write
$K_{BFKL}~=~c_1~ln^3~R ~+~c_2~[ln^4~R ~-~{\cal K}_2]^{1\over 2}~,  $
where $ln^3~R$ is antisymmetric under $1 \leftrightarrow 2$ and 
$[ln^4~R ~-~{\cal K}_2]^{1\over 2}$ is 
symmetric. ${\cal K}_2$ is defined by the eigenvalue spectrum 
$\Lambda(\nu,n)$. Possibly $K_{BFKL}$ can be usefully defined as a 
holomorphic extension of $ln^3 R$. 

$ln^mR$ is directly related to the diagrams involving
two pairs of points joined by $m$ propagators and would naturally appear in a 
high-order conformal approximation to $K_{BFKL}$. An all-orders sum might 
then produce powers of $R$. Whether $t$-channel unitarity can be used to 
discuss such contributions is, at present, an open question. Understanding
the relationship between $K^{(4)}$ and the exact NLO kernel\cite{vf} should
help. W\"usthoff is currently investigating whether the spectra of $ln^m~R$
can be obtained via the generating function
$\cal{G}(R,\delta)~=~R^{\delta}$. 

\mainhead{4. MASSLESS QUARK REGGEON INTERACTIONS}

In the last Section I described the use of Multi-Regge theory to obtain
perturbative Regge results. At a more ambitious level, I would like 
to use the same formalism to study\cite{arw1} the full dynamical Pomeron and
include (derive?) confinement and chiral symmetry breaking. The essential
idea is that massless quarks give anomalous triple-Regge interactions
(directly related to the triangle anomaly) producing the axial charge
violation normally associated with non-perturbative instanton interactions.
These interactions produce zero quark-mass reggeon diagram infra-red 
divergences. The divergent amplitudes contain both hadrons (as 
multi-quark reggeon states) and the dynamical Pomeron. To carry through a full 
analysis of the divergences and diagrams involved requires maximal use of
all the Multi-Regge theory ingredients listed above together with the
necessary QCD calculations. In the following I will be able to provide
little more than a glimpse of what is involved. 

\noindent \parbox{3.1in} { $~~$  For multiparticle amplitudes, 
reggeon unitarity, is very 
powerful. For example, consider the eight-particle amplitude in the 
``helicity-pole limit''
$ u_1,u_2,u_3,u_4~\to \infty $. The S-W representation determines that only
one (analytically-continued) partial-wave amplitude is involved.
Since this 
} 
\parbox{2.3in}{
\begin{center}
\leavevmode
\epsfxsize=1.7in
\epsffile{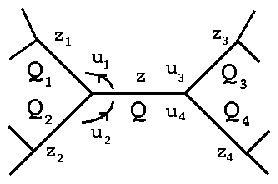}
\end{center}}
\newline 
amplitude satisfies reggeon 
unitarity in all $t$-channels,
the asymptotic 
behavior can be completely represented by transverse momentum integrals of the 
form
\newline \parbox{2.5in}{
\leavevmode
\epsfxsize=2.2in
\epsffile{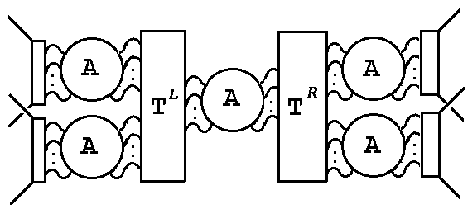}
}
\parbox{2.9in}{ where ${\raisebox{-1mm}{\epsfxsize=0.2in \epsffile{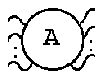}}}$
contains all elastic scattering 
reggeon diagrams. $T^L,T^R$ contain connected and disconnected interactions 
that involve both elastic scattering (helicity non-flip) reggeon
vertices and also new } 
\newline ``helicity-flip'' vertices. The new vertices can
be studied in a ``non-planar'' triple-regge 
limit. Consider three quarks scattering 
via gluon exchange with the triple-gluon
coupling given by a quark loop and take the limit 
\newline \parbox{1.4in}{
\leavevmode
\epsfxsize=1.3in
\epsffile{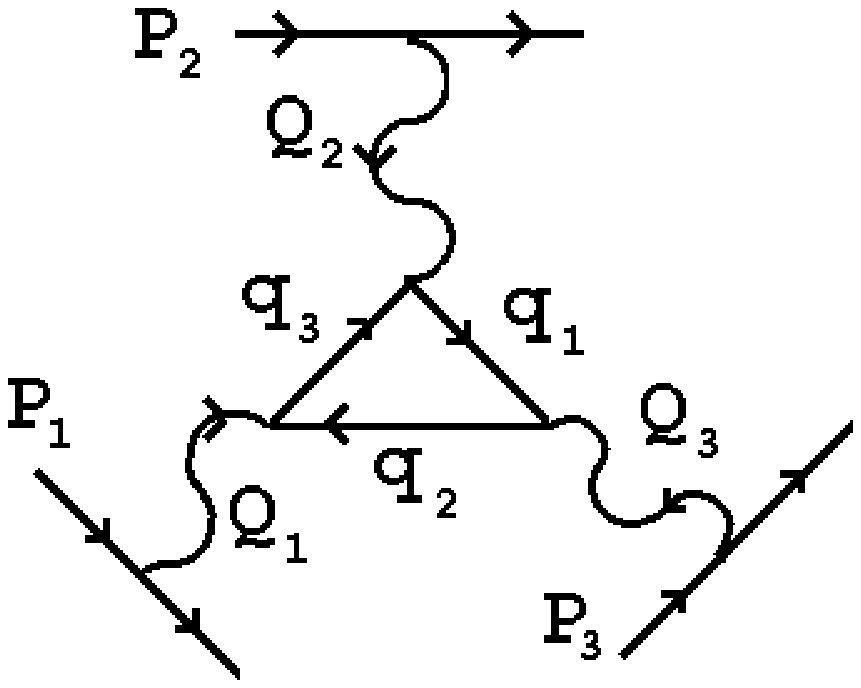}
}
\parbox{4in}{ $P_1 \sim (p_1,p_1,0,0),~
P_2 \sim (p_2,0,p_2,0),~ P_3 \sim  (p_3,0,0,p_3), $
\newline $\forall p_i \to \infty $. The resulting amplitude has the form 
\vspace{0.1in}
\newline $~~~~~~ ~~~~~~~~~~g^6~ { p_1p_2p_3 \over Q^2_1 Q^2_2 Q^2_3 } 
~\Gamma_{1^+2^+3^+}(q_1,q_2,q_3) $
\newline 
where \newline
$~~\Gamma_{\mu_1 \mu_2 \mu_3} = i\int {  d^4 k~ Tr \{ \gamma_{\mu_1}
(\st{q}_3 + \st{k} + m ) \gamma_{\mu_2} (\st{q_1} + \st{k} + m ) 
\gamma_{\mu_3} (\st{q}_2 + \st{k} + m) \} 
\over [ (q_1 + k)^2 - m^2 ][ (q_2 + k)^2 - m^2 ]
[ (q_3 + k)^2 - m^2 ]}$
} 
\vspace{0.1in}
\newline is the usual quark triangle diagram amplitude and 
$\gamma_{i^+} = \gamma_0 + \gamma_i $ . 
Because of the triangle singularity, there is an ``infra-red anomaly'' in 
that the 
limits $q_1, q_2, q_3 \sim Q
\to 0$, $m \to 0$ 
do not commute 
for the $O(m^2)$ part of 
$\Gamma_{1^+2^+3^+}$, i.e.  

$~~~~~~~~~~~~~~~~~~~\Gamma_{1^+2^+3^+,m^2}
{\centerunder{$\large\sim$}{\raisebox{-3mm} 
{$\scriptstyle Q \to 0$} }}~Q ~i~m^2 \int {d^4k \over [ k^2 - m^2 ]^3 }
 ~~{\centerunder{$\longrightarrow $}{\raisebox{-3mm} {$\scriptstyle
m \to 0 $}} }~ ~  R ~Q $

Including color factors and summing diagrams we find 
that $\Gamma_{1^+2^+3^+,m^2}$ appears only in those 
triple-regge vertices where all reggeon states have ``anomalous color parity''
(i.e. color parity $\neq$ signature) e.g. 
\newline
\parbox{1.6in}{
\leavevmode
\epsfxsize=1.4in
\epsffile{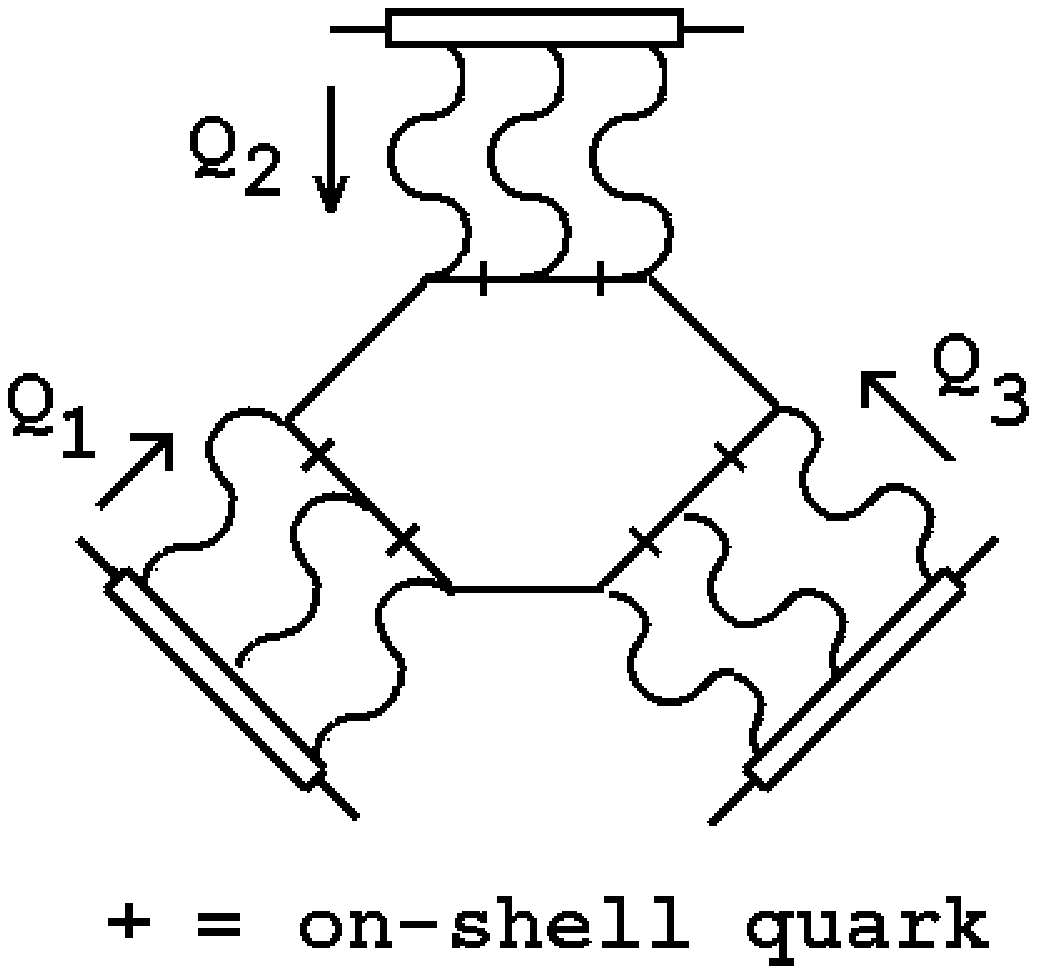}
}
\parbox{3.8in}{ In this diagram each three reggeon state has odd signature 
but the color factor must contain a product of $f$ and $d$-tensors that
makes the color parity even (c.f. the winding-number current
$K_i^{\mu}= \epsilon^{\mu \nu \gamma 
\delta}f_{ijk}d_{jrs}A^k_{\nu}A^r_{\gamma}A^s_{\delta}$). In such vertices
the ``anomalous'' survival of $O(m^2)$ quark helicity-flip processes, when
$Q_i \to 0$ before $m \to 0$, potentially reproduces the axial charge
violation of 
instanton interactions. The presence of} 
\newline $\Gamma_{1^+2^+3^+,m^2}$ leads to the
breakdown of the usual Ward identity cancellations
when $m \to 0$ in 
(non-planar) multi-regge diagrams where $Q_1 \sim
Q_2 \sim Q_3 \sim 0$ is part of the integration region. As a result an 
infra-red divergence appears, e.g. in the diagram 
\vspace{0.1in}
\newline \parbox{2.3in}{
\leavevmode
\epsfxsize=2in 
\epsffile{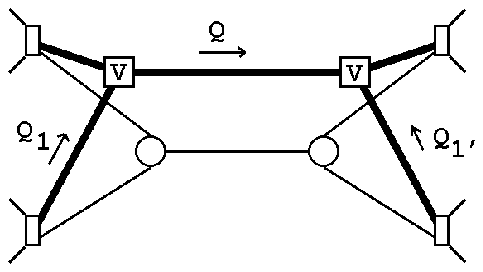}
}
\parbox{3.1in}{
a divergence will occur for $Q,Q_1,Q_{1'} \sim 0$ as $m \to 0$ if $V$ is
anomalous, i.e. if \epsfxsize=0.15in \epsffile{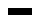} ~is an anomalous
color parity reggeon state. To show that this divergence is not cancelled
by other diagrams requires a systematic analysis in which
the SU(3) gauge symmetry is
 } 
\newline initially broken to SU(2) (c.f. an 
instanton interaction is always associated with SU(2) subgroup).
With the SU(3) symmetry broken, a divergence occurs when 
\epsfxsize=0.2in \epsffile{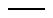} ~~is an SU(2) singlet state containing 
one or more massive gluons (or quarks) and 
\epsfxsize=0.15in \epsffile{dss15.ps} ~carries anomalous color parity. 

Divergent higher-order diagrams involve $V$ vertices in a similar manner e.g.  
\newline \parbox{2.5in}{ 
\leavevmode
\epsfxsize=2.2in
\epsffile{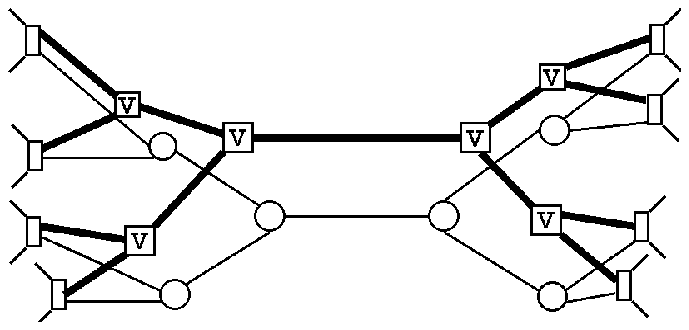}
}
\parbox{2.9in}{ 
An overall logarithmic divergence occurs 
when all $Q_i$ entering each $V$ vertex vanish.
The coefficient of the divergence gives ``physical amplitudes'' in which
all anomalous reggeon states  carry zero $k_{\perp}$. Effectively physical 
reggeon states are SU(2) singlet ``parton'' states in a }
\newline background ``reggeon condensate''. Both hadrons and the Pomeron can
be described this way. I will not give details here. However, I am optimistic 
that I have finally resolved\cite{arw1} how
the ``Super-Critical Pomeron'' of Reggeon Field Theory is realized in QCD
and that many other attractive results follow. 

{\it  In first approximation, (in the condensate 
background) the Pomeron is a reggeized
gluon and hadrons are ``constituent quark'' reggeon states
 with the confinement and chiral symmetry breaking spectrum. 
The Critical Pomeron is inter-related with the 
restoration of SU(3) gauge symmetry ...}

\end{document}